\documentstyle[psfig]{mn}

\input{epsf}

\begin{document}
\pagestyle{headings}


%
%
\newcommand{\etal}{{\rm et al.}~}
\newcommand{\hmpc}{{\rm h$^{-1}$Mpc}}
\newcommand{\mhmpc}{{\, \rm h^{-1}Mpc}}
\newcommand{\beq}{\begin{equation}}
\newcommand{\eeq}{\end{equation}}
\newcommand{\spr}{\marginpar{\sc Check!}}
\newcommand{\mcom}[1]{\marginpar{\sc #1}}
%
%
\newcommand{\eg}{{e.g.$\,$}}   

\newcommand{\delg}{\delta^{\rm gal.}}
\newcommand{\ste}{\sigma_{{\de}\vert{\te}}}
\newcommand{\sde}{\sigma_{{\te}\vert{\de}}}
\newcommand{\der}{{\rm d}}
\newcommand{\qniec}{\end{document}}
\newcommand{\de}{\delta}
\newcommand{\te}{\theta}
\newcommand{\varte}{\vartheta}
\newcommand{\veps}{\varepsilon}
\newcommand{\Sig}{\Sigma}
\newcommand{\lam}{\lambda}
\newcommand{\p}{\partial}
\newcommand{\f}{\frac}
\newcommand{\s}{\sigma}
\newcommand{\bfx}{{\mathbf x}}
\newcommand{\bfr}{{\mathbf r}}
\newcommand{\bfs}{{\mathbf s}}
\newcommand{\bft}{{\mathbf t}}
\newcommand{\bfz}{{\mathbf z}}
\newcommand{\bfy}{{\mathbf y}}
\newcommand{\bfk}{{\mathbf k}}
\newcommand{\bfv}{{\mathbf v}}
\newcommand{\bfq}{{\mathbf q}}
\newcommand{\bfg}{{\mathbf g}}
\newcommand{\bfp}{{\mathbf p}}
\newcommand{\bfu}{{\mathbf u}}
\newcommand{\calD}{{\mathcal D}}
\newcommand{\calF}{{\mathcal F}}
\newcommand{\calO}{{\mathcal O}}
\newcommand{\calQ}{{\mathcal Q}}
\newcommand{\calC}{{\mathcal C}}
\newcommand{\calI}{{\mathcal I}}
\newcommand{\calL}{{\mathcal L}}
\newcommand{\calK}{{\mathcal K}}
\newcommand{\calN}{{\mathcal N}}
\newcommand{\calS}{{\mathcal S}}
\newcommand{\calH}{{\mathcal H}}
\newcommand{\calP}{{\mathcal P}}
\newcommand{\eps}{{\epsilon}}
\newcommand{\bc}{\begin{center}}
\newcommand{\be}{\begin{equation}}
\newcommand{\ee}{\end{equation}}
\newcommand{\ba}{\begin{eqnarray}}
\newcommand{\ea}{\end{eqnarray}}
\newcommand{\ec}{\end{center}}
\newcommand{\lan}{\langle}
\newcommand{\ran}{\rangle}
\newcommand{\ov}{\overline}
\newcommand{\kms}{{$\mathrm{km}\; \mathrm{s}^{-1}$}}
\newcommand{\cl}{C{\L}97}
\newcommand{\inv}{C{\L}PN}

\title{
Reconstructing Cosmic Peculiar Velocities from the Mildly Nonlinear Density Field}
\author[A.~Kudlicki et al.]{Andrzej Kudlicki, Micha\l\ Chodorowski, Tomasz Plewa and Micha\l\ R\'o\.zyczka\\
        N. Copernicus Astronomical Center, Bartycka~18, 00-716 Warsaw, Poland
        }
\maketitle

\begin{abstract}
We present a numerical study of the cosmic density vs.\ velocity divergence 
relation (DVDR) in
the mildly non-linear regime. 
We approximate
the dark matter as a non-relativistic pressureless fluid, and solve its
equations of motion on a grid fixed in comoving coordinates.
Unlike N-body schemes, this method yields directly the volume-averaged 
velocity field. The results of our simulations  are compared with 
the predictions of the third-order perturbation
theory (3PT) for the DVDR. We investigate both the mean
`forward' relation (density in terms of velocity divergence) and the mean
`inverse' relation (velocity divergence in terms of density), with
emphasis on the latter. On scales larger than about 20 megaparsecs,
our code recovers the predictions of 3PT remarkably well,
significantly better than recent N-body simulations.
On scales of a few megaparsecs, the DVDR predicted by 3PT differs
slightly from the simulated one. 
In particular, approximating the inverse DVDR by a 
third-order polynomial turns out to be a poor fit.
We propose a simple analytical description of the inverse relation,
which works well for 
mildly non-linear scales.
\end{abstract}

\section{Introduction} 
\label{sec:intro}
%

It is now widely believed that the large-scale structure formed by the
growth of small inhomogeneities present in the early
Universe. In this scenario, commonly referred to as the 
{\em gravitational instability}
(GI) paradigm, cosmic density and velocity fields are tightly coupled,
and the relation between them involves the
cosmological parameter $\Omega$. In the linear regime, i.e. for 
the r.m.s.\ density fluctuations much smaller than unity, the
density--velocity divergence relation (DVDR) reduces to

\be \de(\bfr) = - f^{-1}(\Omega,\Lambda) \nabla \cdot \bfv(\bfr) \,.
\label{eq:lin}
\ee 
Here, $\de$ is the mass density fluctuation field, $\bfv$ is the
peculiar velocity field, distances are expressed in units of \kms, and

\be
f(\Omega,\Lambda) \simeq \Omega^{0.6} + \f{\Lambda}{70} \left(1 +
\f{\Omega}{2}\right)
\label{eq:f_factor}
\ee
(Lahav \etal 1991). The factor $f$ depends mainly on $\Omega$ and only
weakly on the cosmological constant $\Lambda$ (provided that $\Lambda$
is in the range allowed by observations). The comparisons
between density and velocity fields are a useful test of
the GI hypothesis. In principle, they may also be used as a tool 
to measure $\Omega$ (Dekel \etal 1993).

However, there is both theoretical (e.g., Kaiser 1984;
Davis \etal 1985; Bardeen \etal 1986; Dekel \& Silk 1986; Cen \&
Ostriker 1992; Kauffmann, Nusser \& Steinmetz 1997; Blanton \etal
1998; Dekel \& Lahav 1998) and observational (e.g., Davis \&
Geller 1976; Dressler 1980; Giovanelli, Haynes \& Chincarini 1986;
Santiago \& Strauss 1992; Loveday \etal 1996; Hermit \etal 1996; Guzzo
\etal 1997; Giavalisco \etal 1998; Tegmark \& Bromley 1998)
evidence that
galaxies are biased tracers of the matter distribution. As a result,
the comparisons between the fields in question within linear theory
cannot yield an estimate of $\Omega$ itself. What is actually
measured is the quantity $\beta \equiv \Omega^{0.6}/b$, where $b$ is
the linear bias parameter.

The current state of estimates of $\beta$ is confused. The so-called
velocity--velocity comparisons generally result in low values of
$\beta$ ($\simeq 0.5$: Roth 1994; Schlegel 1995; Schaya, Peebles \&
Tully 1995; Davis, Nusser \& Willick 1996; da~Costa \etal 1997; Riess
\etal 1997; Willick \etal 1997; Willick \& Strauss 1998), while
density--density comparisons yield high values ($\simeq 1.0$: Dekel
\etal 1993; Hudson \etal 1995; Sigad \etal 1998). In
velocity--velocity comparisons, galaxy density field is used to
predict the associated peculiar velocity field, which in turn is
compared to the observed peculiar velocities of a sample of galaxies
with measured redshift-independent distances. In density--density
comparisons, velocity data are used to reconstruct the underlying mass
density field, in order to compare it with an observed galaxy density
field. A number of possible explanations of the divergence in the
estimated values of $\beta$ has been proposed (see, e.g., Sigad \etal
1998). One of them are non-linear effects.

The density fluctuations obtained from current redshift surveys
(e.g. Fisher \etal 1994) and from the {\sc potent} (Dekel \etal 1998)
reconstruction of the mass density field slightly exceed the regime of
applicability of linear theory. For example, the density contrast in
regions like the Great Attractor or Perseus-Pisces is around unity
even when smoothed over scales of $1200$ \kms, currently employed in
density--density comparisons (Sigad \etal 1998). In velocity--velocity
comparisons, the fields in question are generally smoothed over
smaller scales than in density--density ones. Astonishingly, while in
current density--density comparisons the non-linear corrections to the
linear density--velocity relation, equation~(\ref{eq:lin}), 
are accounted for, in
velocity--velocity comparisons they are not. The only exception is an
attempt by Willick \etal (1997) to model the DVDR by a second-order
formula.\footnote{Strictly speaking, they  proposed a
fully non-linear formula, but in the process of actual comparison they
truncated it at second order terms.} To their surprise,
the maximum-likelihood fit of the predicted to the observed peculiar
velocities was for zero amplitude of the second-order
corrective term. However, the smoothing scale they used was $3
\mhmpc$.  At such a small scale, the variance of the density
field is already in excess of unity and, as we will show later,
neither the linear nor the second-order formula is a good description
of the actual DVDR.

The purpose of this paper is to propose a simple and accurate
description of the DVDR at mildly non-linear\footnote{We define
mildly non-linear scales as these at which the r.m.s.\ density
fluctuation is a significant fraction of, but still smaller than,
unity. Then the mildly non-linear scales in the Universe are about or
greater than 8 \hmpc\ for top-hat smoothing, and roughly twice smaller
for Gaussian smoothing.} scales, which would be easy to implement 
in current
velocity--velocity comparisons. To date, there have been several
attempts to construct a mildly non-linear extension of
relation~(\ref{eq:lin}). They were either based on various analytical
approximations to non-linear dynamics (Bernardeau 1992; Catelan \etal
1995; Chodorowski 1997; Chodorowski \& {\L}okas 1997, hereafter C\L 97; 
Chodorowski \etal 1998, hereafter C\L PN), 
or N-body simulations (Mancinelli \etal 1994; Ganon et
al., in preparation), or both (Nusser \etal 1991; Gramann 1993;
Mancinelli \& Yahil 1995). So far, the most comprehensive description
of the mildly non-linear DVDR has been recently done by Bernardeau \etal
(1999; hereafter B99). Our work is an extension and improvement of B99
in several ways:

\begin{itemize}

\item In B99 the analysis was for technical reasons performed 
solely for 
fields smoothed with a top-hat filter. Here we also analyze fields
smoothed with a Gaussian filter, which is now commonly applied
to observational data.

\item The fully non-linear formula proposed by B99 expresses density in
terms of the velocity divergence (the so-called `forward'
relation). However, in velocity--velocity comparisons one needs a
formula for the velocity (divergence) expressed as a function of the
density ( the so-called `inverse' relation). Due to the scatter in the
DVR, the latter is not given by a straightforward inversion of
the former. We obtain such an `inverse' formula here.

\item Our `inverse' formula is much simpler compared to the `forward'
one of B99, but equally  accurate, as detailed comparisons with
numerical simulations show. Unlike the second order formula 
used by Willick \etal (1997), 
it works well for smoothing scales down to a few megaparsecs.

\item Instead of performing N-body simulations, we 
model cold dark matter as a pressureless cosmic
fluid. We solve non-linear 
equations for its evolution
on a grid fixed in comoving coordinates. 
This approach is advantageous over the standard N-body
one for studying the evolution of the {\em velocity\/} field in 
the mildly non-linear regime.
The reasons are outlined
below.

\end{itemize}

Both in N-body simulations and in our code, the final velocity field
is known at a discrete set of points. In the case of an N-body
simulation this set is particles' positions, in our case it is  the
grid. Due to clustering, the N-body velocity field is sampled very
non-uniformly, while the sampling of our velocity field is
perfectly uniform. Smoothing of a non-uniformly sampled velocity field
leads to the so-called `sampling gradient bias' (Dekel, Bertschinger
\& Faber 1990). In N-body simulations, the sampling rate of the
velocity field is proportional to the number density of particles in a
given region. The averaging of the field within a smoothing window is
therefore not volume- but mass-weighted, resulting in a special 
type of bias mentioned above. To circumvent this problem, elaborate 
`tessalation' algorithms
for the velocity field have been proposed (Bernardeau \& van de
Weygaert 1996). However, they  work only for a top-hat filter. Another
problem is that N-body simulations provide very little information on
the velocity field in voids, simply because there are very few
velocity tracers there.

Due to uniform sampling, our simulations yield directly
volume-weighted values of velocity, for any type of smoothing.
Moreover, we probe the velocity field in the voids as 
finely as in dense regions. As a result, at a very low
numerical cost it was possible to have the velocity field sampled at a
comparable number of points to that of B99 ($64^3$ compared to
$50^3$), and still of significantly better quality, as shown below.

The paper is organized as follows: 
In 
Section~\ref{sec:theo} 
we discuss the theoretical 
aspects of the DVDR. Then, in
Section~\ref{sec:simu}, we present our simulations, describing
the algorithm in
Section~\ref{sec:code}
and the cosmological model investigated in
Section~\ref{sec:seru}.
In Section~\ref{sec:anco} we investigate the
mean forward density--velocity relation and we
demonstrate that it is well described by a 
third-order polynomial.
In Section~\ref{sec:rnco} we show that the  polynomial formula
is a poor approximation of the {\em inverse\/} relation, and we propose
an alternative description in
Section~\ref{sec:alfa}.
We summarize our results in 
Section~\ref{sec:suma}.

\section{Theoretical framework}
\label{sec:theo}
Due to the Kelvin circulation theorem, 
the cosmic velocity field remains irrotational
before shell-crossings. 
It can therefore be
described by a single scalar function, which we choose here to be the
velocity divergence, 
$\nabla \cdot \bfv$ (throughout the paper the derivative is taken 
in velocity units, i.e. $H=1$).
The linear
relation~(eq.~\ref{eq:lin}) between the density contrast and the velocity
divergence at a given point holds only on scales large enough so that
the density fluctuations are much smaller compared to unity. On
smaller scales, non-linear effects modify the relation in a number of
ways. For full discussion of the density versus velocity divergence
relation in the mildly non-linear regime the reader is referred
to B99. 

In brief, qualitative features of the relation can be outlined
as follows:

\begin{itemize}

	\item It is non-linear.

	\item It is also non-local, which implies that it
        is locally non-deterministic, i.e. it has a scatter  
        in $\de$ for a given $\nabla \cdot \bfv$ and vice versa.

	\item Since the scatter originates exclusively from higher 
        (than linear) order terms, it is small in the mildly
        non-linear regime. Therefore, the most probable values of
	$\de$ and $\nabla \cdot \bfv$ form an elongated region in the
	$(\de,\nabla \cdot \bfv)$ plane.

\end{itemize}

\begin{figure}
\vspace{-0mm}
\centerline{
\hspace*{0mm}
\hfill
\epsfxsize=80mm\epsffile{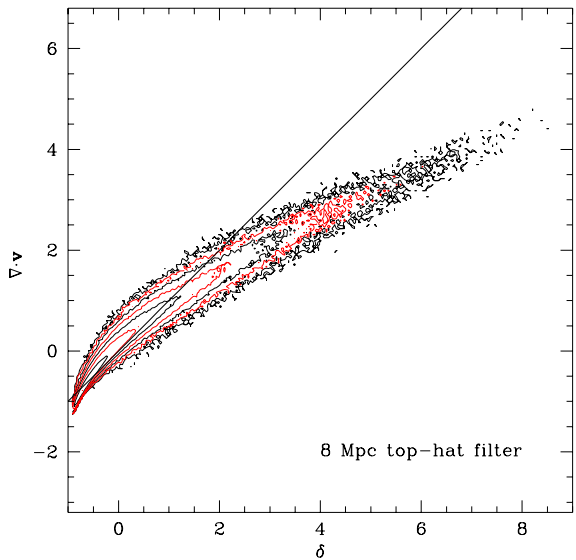}
\hfill
           }
\vspace{-0mm}
\caption{
Joint probability distribution function of density and velocity
divergence, $P(\de , \nabla \cdot \bfv)$, from a $64^3$ 
simulation. Data are convolved with a 8\hmpc\ top-hat filter.
Contours
have intervals of 0.5 in $\log_{10}(P)$. Solid line represents
the linear relation.
\label{fig:bananth}}
\end{figure}

Figure~\ref{fig:bananth} is a typical plot of the values of $\de$ and
$\nabla \cdot \bfv$ obtained in our simulations. The fields are smoothed with a
top-hat filter with the smoothing radius of 8 \hmpc. For such a
smoothing scale, the r.m.s. fluctuation of the density field,
$\s_{\de}$, in our simulations is $\simeq 0.9$, so the fields are
close to leave the regime of mild non-linearities. However, in 
Figure~\ref{fig:bananth} one can still observe an obvious 
correlation between the  density and velocity divergence.

Analytical calculations predict (Bernardeau 1992; Gramann 1993; 
Catelan \etal 1995; Mancinelli \& Yahil 1995; Chodorowski 1997; 
C\L 97; C\L PN) and
N-body numerical simulations confirm (Mancinelli \etal 1994; B99; 
Ganon \etal 1999) that the mildly non-linear DVDR depends on $\Omega$ 
and $\Lambda$ in a very simple way. Specifically, if we define the 
{\em scaled\/} velocity divergence,

\be
\te \equiv - f^{-1}(\Omega,\Lambda) \nabla \cdot \bfv \,, 
\label{eq:scaled}
\ee 
the relation between the density and the scaled divergence (for
simplicity it will also be referred to as DVDR) will be practically 
$\Omega$- and
$\Lambda$-independent. Since the relation has a scatter, the full
information about the DVDR is contained in the joint probability
distribution function (PDF) for $\de$ and $\te$. 
Such a joint PDF has been constructed by B99.
However, the scatter is small compared to
random errors in the observed density and velocity fields
(\inv). Therefore, of most interest for practical applications are the
mean relations: the mean density for the given velocity divergence, $\lan
\de_{|\te}\ran$ (the `forward' relation), and vice-versa, $\lan \te
_{|\de}\ran$ (the `inverse' relation).\footnote{Due to the scatter,
the inverse relation is not given by a straightforward inversion of
the forward one.} The forward relation is relevant for
density--density comparisons; the inverse relation is relevant for
velocity--velocity comparisons. Since velocity--velocity
comparisons employ smaller smoothing lengths, non-linear effects are
more important there than in density--density comparisons. That is why
in this paper we shall concentrate on finding a simple, and
simultaneously robust, description of the inverse relation for
Gaussian smoothing of the fields. 

Though one might formally derive the inverse relation from the joint
PDF constructed by B99, it would be inappropriate for a number of
reasons. Firstly, while the forward relation can be derived from this
PDF in an analytic form, the inverse one can only be computed
numerically. Secondly, the joint PDF was constructed by B99 for
top-hat smoothed fields and it is expected to depend quantitatively on
the type of smoothing. Finally, with our fluid code we hope
to trace the actual DVDR more accurately.

The mildly non-linear regime is the one in which perturbation theory
can be applied. In particular, the mean relations are a priori
accessible to analytical perturbative calculations. \cl\ derived the
forward DVDR up to third-order terms, accounting for the smoothing of
the density and velocity fields. The mean density contrast given the
scaled velocity divergence is a third-order polynomial in the
divergence,
\be 
\langle\de_{|\te}\rangle = 
a_0 + a_1 \te + a_2 \te^2 + a_3 \te^3 \,,
\label{eq:for}
\ee 
where $a_0 = - a_2\s_\te^2$ and $\s_\te^2 $ 
is the variance of the 
scaled velocity divergence
field,
$\langle\te^2\rangle$.
The coefficients, $a_i$, appearing in the above expansion were
explicitly calculated by \cl\ for Gaussian smoothing and by B99 for
top-hat smoothing. As explained above, they depend extremely weakly on
$\Omega$ and $\Lambda$. \inv\ derived the inverse relation
up to third-order terms,
\be
\langle\te_{|\de}\rangle = 
r_0 + r_1 \de + r_2 \de^2 + r_3 \de^3 
\,. 
\label{eq:inv}
\ee
The coefficients $r_i$ were calculated by \inv\ for Gaussian smoothing
and by B99 for top-hat smoothing.

Contributions to the DVDR from orders higher than third are known in
only one special case, of unsmoothed fields with vanishing variance.
Bernardeau (1992) derived for this case the following formula:
\be
\langle\te_{|\de}\rangle = 
\f{3}{2} \left[ (1 + \de)^{2/3} - 1 \right] 
\,.
\label{eq:b92}
\ee
The above expression is strictly valid only for $\s_\de \to 0$ {\em
and\/} $\Omega \to 0$, but since the $\Omega$-dependence of the
(scaled) DVDR is extremely weak, it remains a good approximation also
for other values of $\Omega$. 

Equations~(\ref{eq:inv}) and~(\ref{eq:b92}) are two different
approximations to the inverse relation. As already stated,
equation~(\ref{eq:inv}) accounts for smoothing and for finite 
variances of the fields. 
Equation~(\ref{eq:b92}) does not,\footnote{B99 argued
however that this result should remain valid for top-hat smoothed
fields with vanishing variance.} but instead it includes contributions
from all orders. We can therefore expect the two equations to carry 
complementary information about the actual relation. Our procedure of
finding a simple and accurate description of the inverse relation will
consist of two steps. Firstly, we will check on which scales the
third-order expression~(\ref{eq:inv}) is a good description of the
relation, and at which scales it already fails. Then, guided by our
numerical results, and by equation~(\ref{eq:b92}), we will
look for a formula for the inverse relation, which
would be accurate in the whole range of the mildly non-linear scales.

\begin{figure}
\vspace{-0mm}
\centerline{
\hspace*{0mm}
\hfill
\epsfxsize=80mm\epsffile{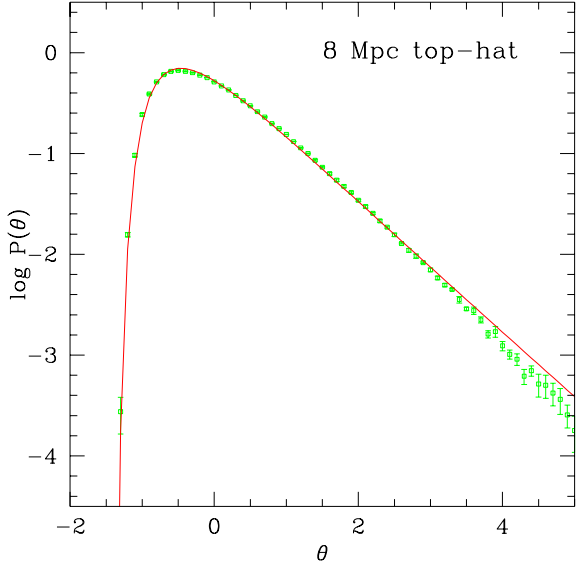}
\hfill
           }
\vspace{-0mm}
\caption{$P(\theta)$, the probability distribution of 
the scaled velocity divergence
for 8\hmpc\ top-hat smoothing. Open squares are combined results
from our six runs, with error-bars shown. Solid line represents
formula (12) of Bernardeau (1994), with the variance of $\theta$ 
taken to be the average from our simulations. 
\label{pdvt}}
\end{figure}

\begin{figure*}
\vspace{-0mm}
\centerline{
\hspace*{0mm}
\hfill
\epsfxsize=72mm\epsffile{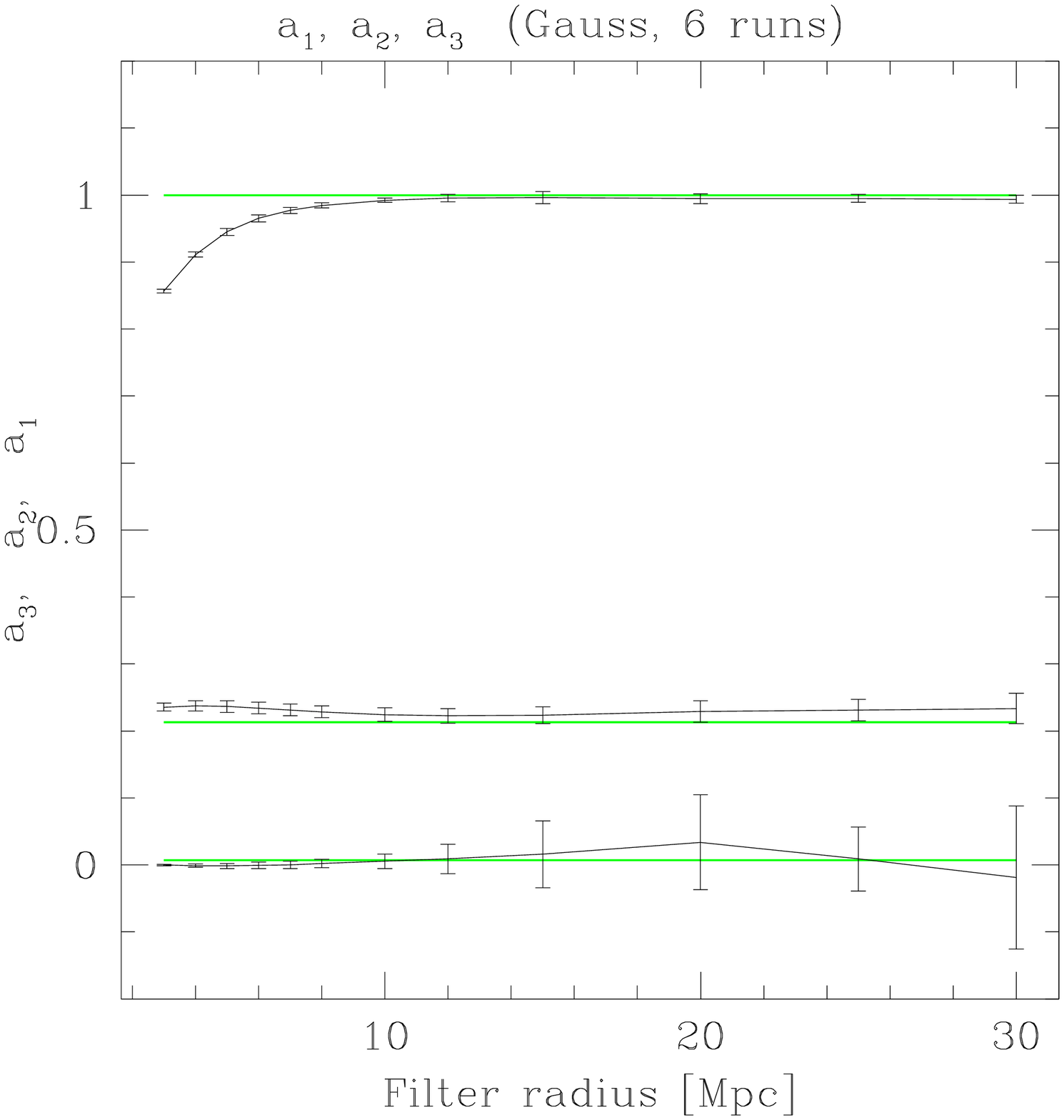}
\hspace*{30pt}
\hfill
\epsfxsize=72mm\epsffile{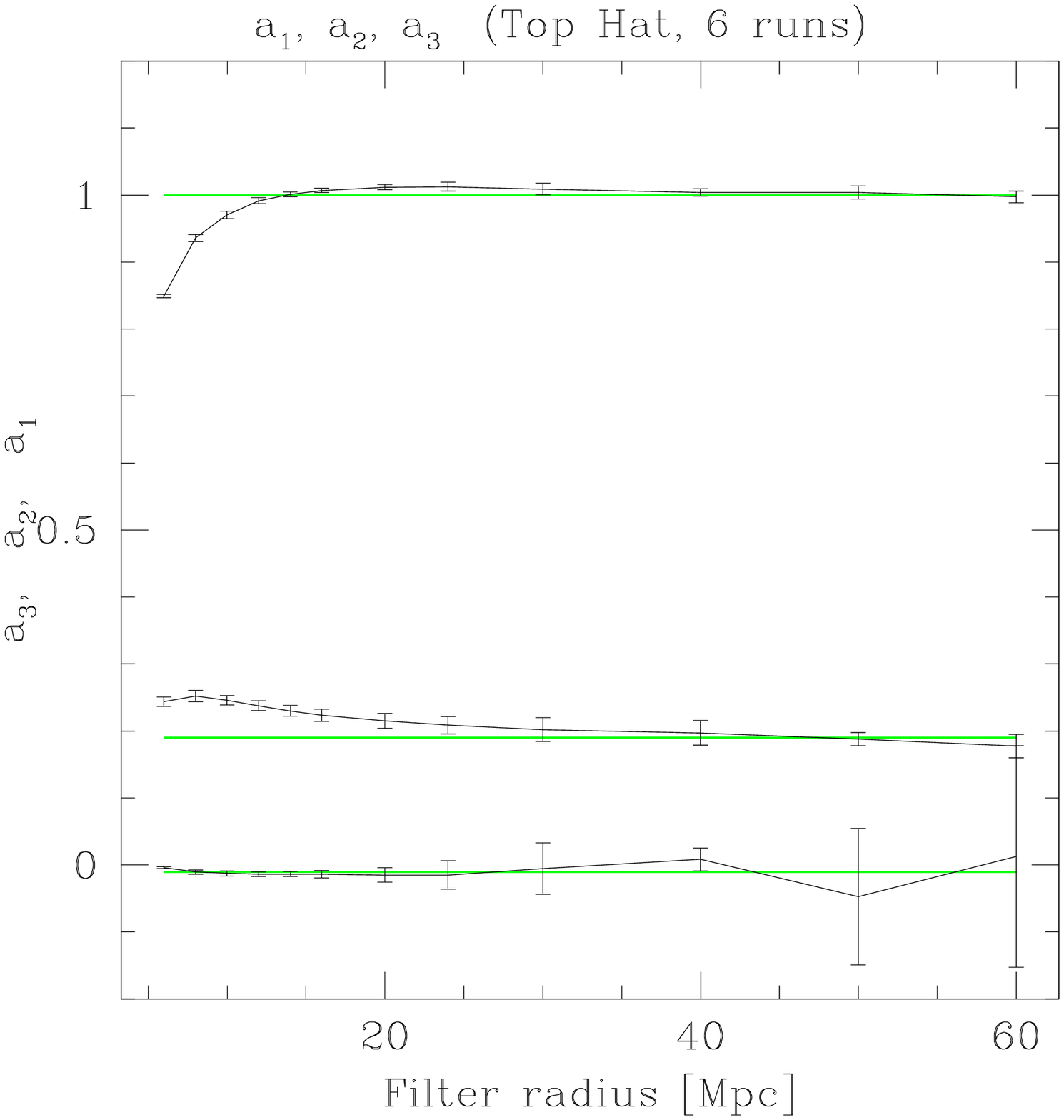}
\hfill
           }
\vspace{-0mm}
\caption{
The coefficients $a_1, a_2, a_3$ from six simulations (curves and error-bars) 
and their third-order perturbation theory predictions (solid lines). 
Left panel: Gaussian filter, right panel: Top-Hat  filter.
\label{antg}}
\end{figure*}

\begin{figure*}
\vspace{-0mm}
\centerline{
\hspace*{0mm}
\hfill
\epsfxsize=72mm\epsffile{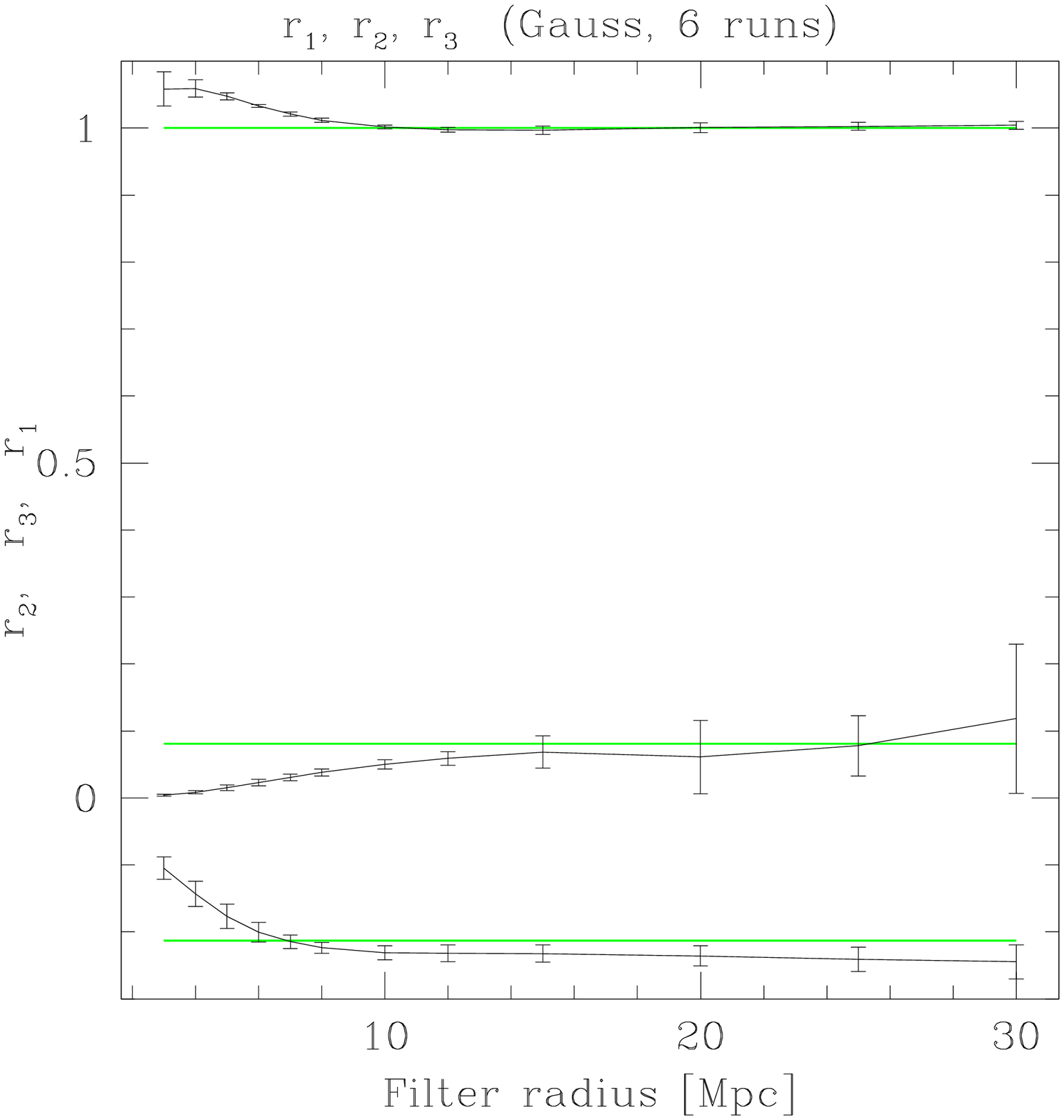}
\hspace*{30pt}
\hfill
\epsfxsize=72mm\epsffile{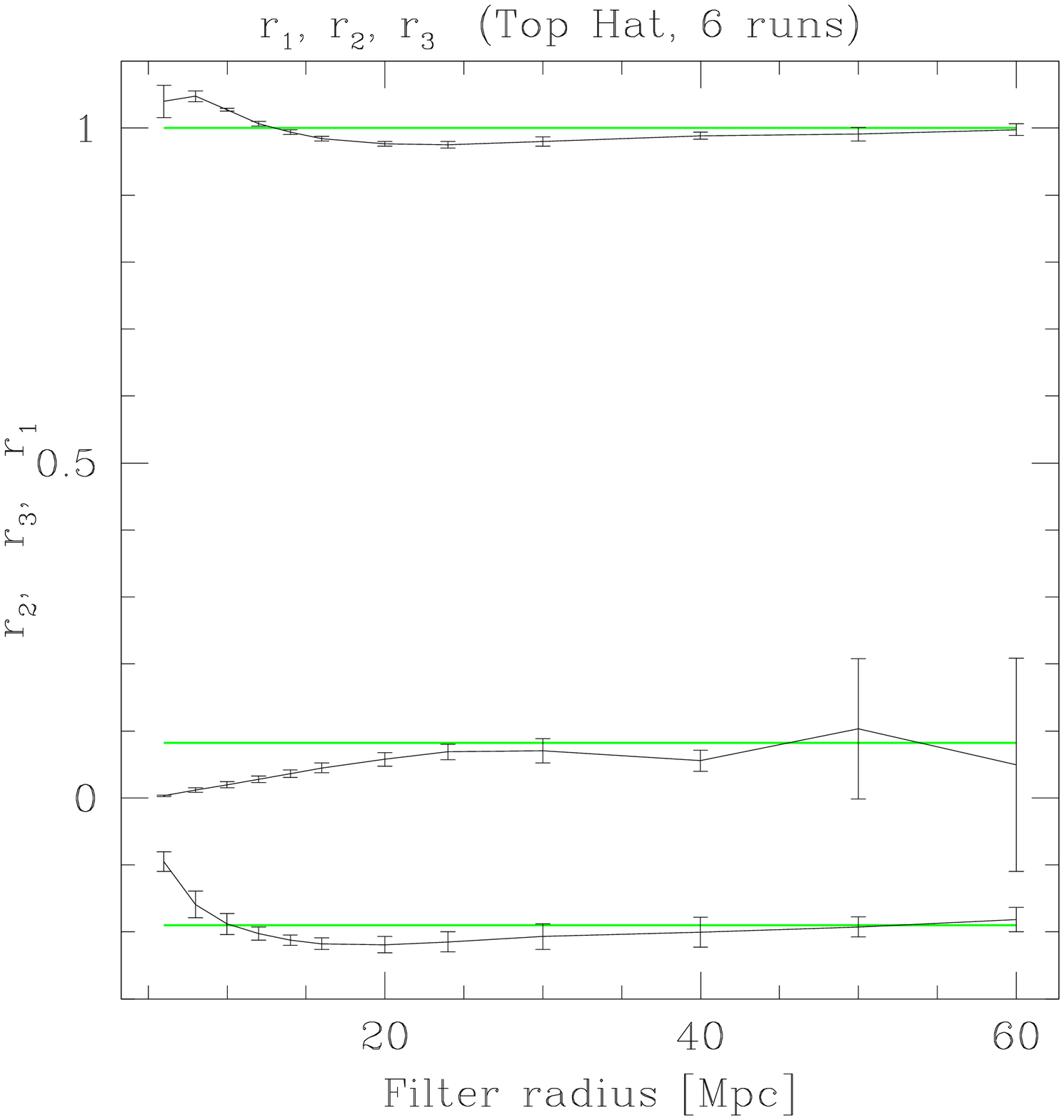}
\hfill
           }
\vspace{-0mm}
\caption{
The coefficients $r_1, r_2, r_3$ from six simulations (curves and error-bars) 
and their third-order perturbation theory predictions (solid lines). 
Left panel: Gaussian filter, right panel: Top-Hat filter.
\label{rntg}}
\end{figure*}

\section{The simulations}
\label{sec:simu}
\subsection{The code}
\label{sec:code}
As stated in Section \ref{sec:intro},
despite their simplicity and numerous advantages, N-body
simulations of cosmic 
velocity fields have several drawbacks. 

To cope with them, we have performed our simulations using
{\em CPPA} (Cosmological Pressureless Parabolic Advection): 
our original Eulerian, uniform-grid based code for
self-gravitating pressureless fluid evolution in an expanding
Universe.  The main ideas of the algorithm are similar to 
those of Peebles (1987), with several modifications.
%
%
%
%
%
An early version of the code is described in
Kudlicki, Plewa \& R\'o\.zyczka (1996); 
later improvements to the code and test results 
will be described in detail in
Kudlicki \etal (in preparation).
CPPA employs a
three-dimensional Cartesian grid fixed in 
dimensionless comoving coordinates (see Gnedin 1995). 
The Poisson equation is solved by a standard
FFT-based routine, working on the same grid as the Euler
solver.   
Advection of mass and 
momenta is done using 
the piecewise-parabolic scheme, 
as described by Colella \& Woodward (1984).
The advection step consists of  a series of 
sweeps along the main axes of the computational domain. 
Unlike Peebles (1987),
we  use a variable timestep, according to the CFL condition.
To allow for shell crossing on small scales, and inhibit
the unrealistically high densities at cluster centres,
we artificially interchange fluxes
across local directional maxima of the density field.
During a sweep, if a local directional maximum of density 
is encountered, and there is matter falling onto it
from both sides,
then 
the fluxes of density and momentum
calculated at the left and right interface of the maximum density cell
are interchanged.
Performed in all 
directions, it successfully inhibits the non-physical transfer of
power into the smallest scales, while conserving the total momentum.
We are aware that our code does not reproduce small-scale structures
properly,
however for the present purpose -- involving 
window functions larger
than 3 \hmpc\ (Gaussian) and 6 \hmpc\ (Top Hat) --
it is an efficient and satisfactory tool.

\subsection{Selection of the parameters and the models}
\label{sec:seru}

Since the relation between density and scaled velocity
divergence depends very weakly on the background cosmological model
(see the previous section),
we were free to choose the convenient and well-tested 
Einstein--de Sitter model ($\Omega=1, \Lambda=0$). 

To estimate random 
errors, we have
performed six realizations of this cosmological model, each of them
with different random phases of the initial density field.

We aimed at investigating the statistics of density and velocity
fields on both intermediate (several megaparsecs) and large (up to
$60\mhmpc$) scales, so, in order to suppress 
the effects of finite
simulation volume size and improve the statistics we decided to make
the simulation box significantly larger than the largest filter
used. A reasonable solution turned out to be a $(200\mhmpc)^3$ cube
with standard periodic boundary conditions. 
For $64^3$ grid cells, the spatial resolution  is $3.125\mhmpc$,
sufficient for our purposes. As a test we 
performed a simulation with
$128^3$ grid cells for a $(200\mhmpc)^3$ cube and the results
remained in good agreement with those obtained with the coarser grid.

Most of the analytical calculations in the discussed regime
have been done for scale-free, or
power-law, power spectra, $P(k)\propto k^n$. These spectra may seem
artificial, none the less they
are believed to approximate the real power spectrum at
least piecewise over significant ranges of wavelengths. In
particular, in the range of scales $\sim 1$--$20 \mhmpc$, the observed
power spectrum is well approximated by a power law (e.g., Sutherland
\etal 1999; Freudling \etal 1999).
That is why we have decided to use
a power law power spectrum in our simulations.
For convenience, we have picked such normalization of the initial 
density contrast, that the amplitude of the linear growing mode 
measured with a 8\hmpc\ top-hat filter at present epoch is unity,
$(1+z_{\rm initial}) \sigma_{8,{\rm initial}}=1$.  
Since we have chosen a scale-free power
spectrum, the results may be simply rescaled to other normalizations.

Observations suggest that  on mildly non-linear scales 
the effective spectral index $n$ 
lies between $-1$ and $-1.5$
(Baugh \& Efstathiou 1993, 1994;
Fisher \etal 1993; 
Feldman \etal 1994; 
Park \etal 1994;
Lin \etal 1996; 
Sutherland \etal 1999). 
For our simulations we have chosen the value of $n = - 1$,
because one of our code tests was to compare the PDF of the velocity
divergence with the analytical formula of Bernardeau (1994). This
comparison was essential to demonstrate the advantage of our code over
N-body simulations in {\em velocity\/} field studies. 

N-body schemes yield mass-weighted velocity fields, resulting in
spurious velocity gradients. These gradients manifest themselves 
as spurious tails in the PDF of $\te$.
Tesselation techniques, invented by Bernardeau \& van de Weygaert (1996)
to overcome this problem,
are very CPU-time-consuming, and the results 
have much lower resolution than the simulations themselves.
(Commonly $50^3$ compared to $128^3$: Bernardeau \& van de Weygaert 1996; 
Bernardeau \etal 1997; B99).  
In contrast, our code yields directly a volume-weighted
velocity field.
%
\label{sec:pdvt}
Figure~{\ref{pdvt} presents the comparison of the PDF of $\te$ as
recovered from our numerical data with the analytical formula of
Bernardeau (1994). The velocity field is smoothed with a top-hat
filter of the radius of $8 \mhmpc$. Note that the PDF obtained from
the simulations has no spurious tails, whatsoever. The definition of
the scaled divergence is such that negative $\te$ corresponds to
positive divergence, i.e., to the expansion of voids. Note how well
the negative tail of our PDF traces the analytical prediction, on over
three decades of the probability value. (Compare Fig.~6\ of 
Bernardeau~1994.) 
In the extreme part of the positive tail,
the analytical PDF slightly overestimates the measured one. This is to
be expected, since the formula of Bernardeau (1994) is only an
approximate fit to the actual PDF, overestimating the value of
skewness of the distribution ($2$ instead of $1.7$ based on PT). 
Indeed, in our simulations the skewness measured with an 8 \hmpc\ 
top-hat filter has the value of $1.77\pm0.11$. 

\section{The forward relation}
\label{sec:anco}
\label{sec:fwd}

With our models, we have tested  the
polynomial approximation of the mean forward DVDR
(Eq.~\ref{eq:for}).

We compare the coefficients
$a_1, a_2$ and $a_3$ computed from our simulations
to  the corresponding third-order PT values
in Figure \ref{antg}.
These coefficients have been computed from the simulations
using a standard four-parameter%
\footnote{We have found that the value of $a_0$ is perfectly 
consistent with $-a_2 \sigma_\theta^2$, and that a 3-parameter fit
with the  $a_0=-a_2 \sigma_\theta^2$ constraint gives the same
values for $a_1$, $a_2$ and $a_3$.}
least square fit on data from each of the 
$64^3$ grid cells (after smoothing with the filters required). 
We have also performed 
a fit  for all the coefficients $a_0$ through $a_5$,
in this case $a_4$ and $a_5$ were consistent with zero, 
and the values of $a_0$ \ldots $a_3$ did not change 
remarkably between these two fits.
With the exception of the most highly non-linear smoothing
scales, the values of $a_n$  very weakly depend on the
filter size, and  are in good agreement
with the perturbation theory predictions of C{\L}97.

\section{The inverse relation}
\subsection{Polynomial parameterization}
\label{sec:rnco}

\begin{figure}
\vspace{-0mm}
\centerline{
\hspace*{0mm}
\hfill
\epsfxsize=72mm\epsffile{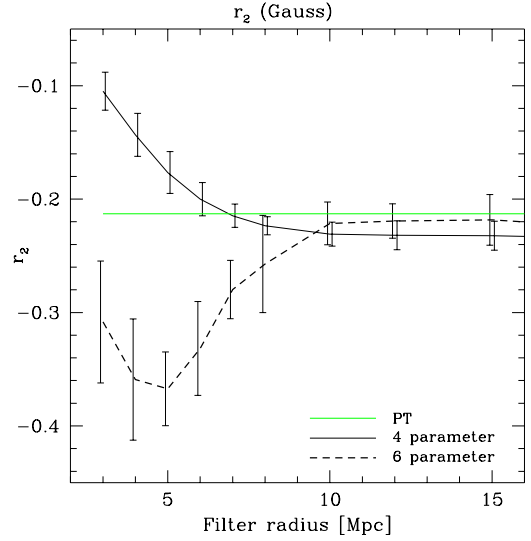}
\hfill
           }
\vspace{-0mm}
\caption{
The value of $r_2$ from 4--parameter fit (thin solid curve) 
and 6--parameter fit (dashed curve). 
Heavy solid line represents the third-order 
perturbation theory value.
\label{r235}}
\end{figure}

As we have shown above, the mean forward relation can be 
described with sufficient accuracy  by the polynomial formula
(\ref{eq:for}) on the relevant scales. 
In this section we test the polynomial
approximation  (\ref{eq:inv}) to the inverse relation.

In Fig.~\ref{rntg} we present our  numerical estimates
of the parameters $r_1$, $r_2$ and $r_3$. 
On scales below $\sim$ 8~Mpc for Gaussian 
and $\sim$ 15~Mpc for top-hat filtering,
their dependence
on the filter size is very strong, especially for
$r_2$, which is the first, and therefore the most essential
parameter describing the non-linearity of the DVDR.
This makes the polynomial formula inconvenient for application to
observational data. Even using the value of  $r_2$ 
predicted for a particular filter size will not help
much since all sizes scale with
the present-epoch density contrast,
$\sigma_{8}$, and the Hubble constant, $h$, 
and neither of these parameters is known accurately yet.

\begin{figure}
\vspace{-0mm}
\centerline{
\hspace*{0mm}
\hfill
\epsfxsize=72mm\epsffile{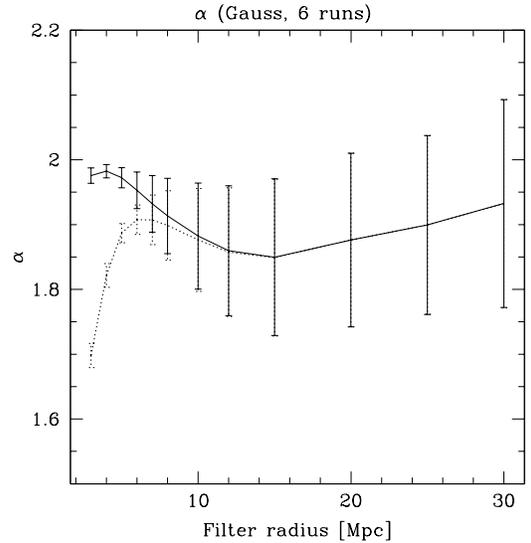}
\hfill
           }
\vspace{-0mm}
\caption{
The $\alpha$ parameter of the $\langle\theta_{|\delta}\rangle$
relation. Dotted curve and error-bars: fit with second-order approximation
to $\epsilon$; 
solid: $\alpha$ from
fit of both $\alpha$ and $\epsilon$.
\label{alph}}
\end{figure}

\begin{figure*}
\vspace{-0mm}
\centerline{
\hspace*{0mm}
\hfill
\epsfxsize=152mm\epsffile{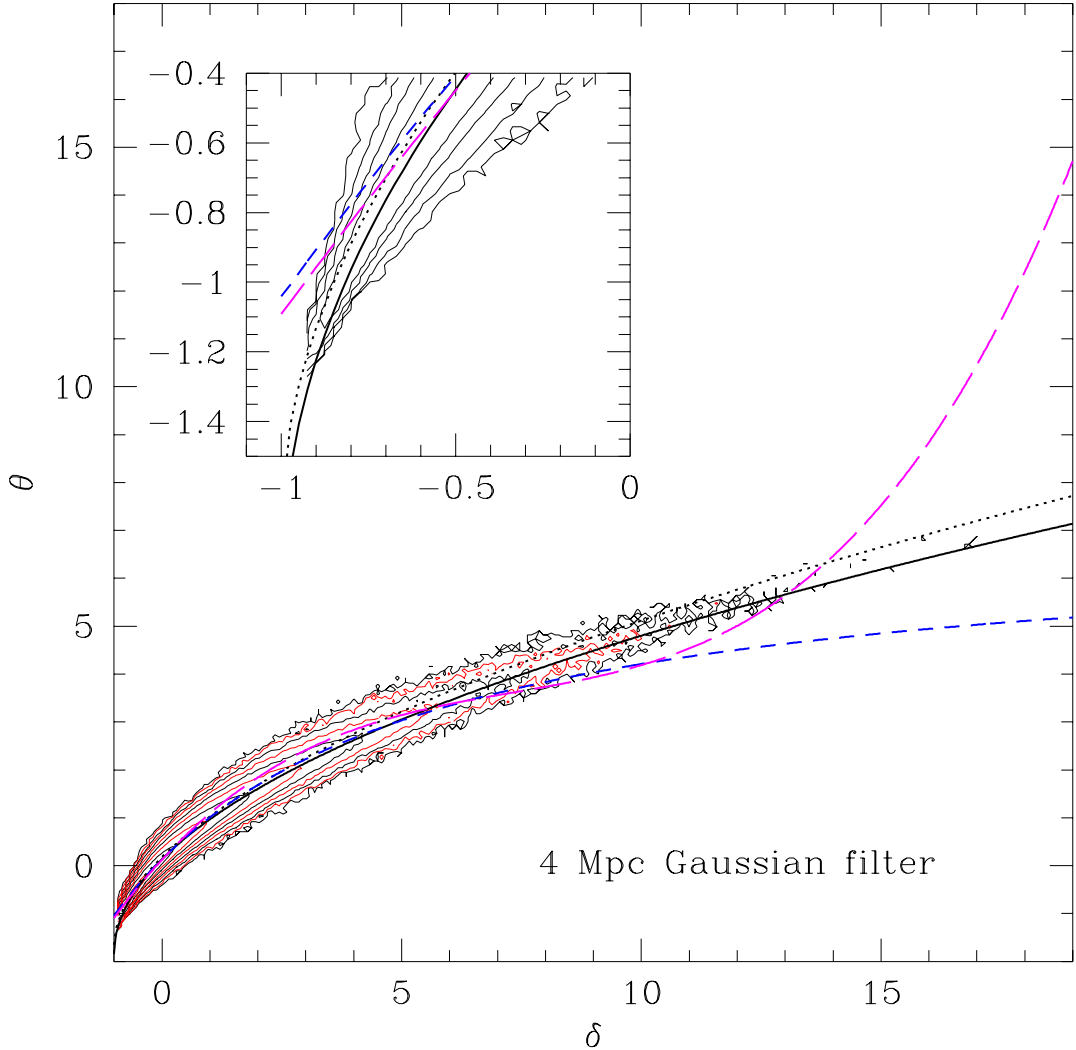}
\hfill
           }
\vspace{-0mm}
\caption{
Joint PDF of $\de$ and $\te$, $P(\de,\te)$: contours
have intervals of 0.5 in $\log_{10}(P)$;
combined data from six simulations are plotted. 
Solid curve represents our formula (\ref{alf}) with 
$\alpha$ and $\epsilon$ fitted independently, 
dotted curve -- a one-parameter fit according 
to formula (\ref{alf}) with the $\epsilon$ offset 
accurate to the second order (eq.~\ref{eps}).
Long-dashed curve represents the third-order polynomial fit
(\ref{eq:inv}), and short-dashed -- the non-linear
formula of Willick \etal (1997). Window in the upper-left
corner enlarges the void region, here for clarity 
contours have intervals of 1.0 in $\log_{10}(P)$.
\label{fig:banangs}}
\end{figure*}
 
We have also performed a fit for the parameters
$ r_0$ through $r_5$ of the 
fifth-order polynomial in $\delta$.
The values
of $r_4$ and $r_5$ turn out to be inconsistent with zero 
at more than $2\sigma$ level for filter sizes smaller than
20 \hmpc,
and, which is still more important, their addition
to the fit significantly influences $r_1$, $r_2$ and $r_3$ 
(see Figure~\ref{r235}).
Moreover, a third-order polynomial that fits the
distribution in the large-scale regime (for small 
$|\delta|$) obviously has a non-monotonic derivative,
which is inconsistent with the properties of the 
actual distribution. 
Figure~\ref{fig:bananth} clearly demonstrates that the
first derivative of the relation is positive and
monotonically decreasing.
The third-order polynomial fit 
to the simulated data is drawn with a long-dashed line 
in Figure~\ref{fig:banangs}.
One can 
observe that this function has an inflection point
well within the range of  $\delta$ and $\theta$ 
occupied by numerical data, not seen 
in the simulated relation.
Another substantial disadvantage of high-order
polynomial fits is the high number of parameters, each of them
depending on the smoothing scale.
Finally, the $r_n$ parameters are strongly correlated,
so possibly another formula, dependent on fewer parameters and 
not a polynomial, will provide a better description.

\subsection{A robust non-linear formula}
\label{sec:alfa}

In search for a better formula for the 
$\langle\theta_{|\delta}\rangle$ relation one needs to
take into account: (a) monotonicity of the fit and its derivative,
(b) agreement with 
the polynomial description 
for large filter radii, (c) proper asymptotic behavior in the 
voids, and (d) the mass conservation law, i.e. the formula
should yield $\langle\theta\rangle=0$.
Of the above, (a), (b) and (c) are satisfied by the asymptotic
formula of Bernardeau (1992), (see eq.~\ref{eq:b92} of
this paper),
which was derived
in the limit of zero density dispersion, 
$\sigma_{\delta} \rightarrow 0$, and which does
not account for smoothing effects.
Filtering of the data
substantially affects the higher order
moments of the distribution of $\de$ and $\te$,
and therefore it is expected to cause a change in the 
shape of the DVDR. 
In order to make equation~(\ref{eq:b92}) applicable to
filtered data (and such are all the data obtained
from galaxy catalogs)
we have made an educated guess, replacing the 
exponent of $2/3$ in the formula with a free parameter, 
$1/\alpha$.%
\footnote{This follows the idea of B99 in a sense, but will result in
a much simpler formula.}
This change does not affect the shape of the fit
in the void wing of the plot, as long as $\alpha$ is not very 
much different from $\frac{3}{2}$.
To satisfy (d), i.e. to keep the average $\theta$ equal zero, 
we add
a constant, depending only on $\alpha$ and scaled by 
the density dispersion. 
Our final formula has the form:
\beq
\theta = \alpha \left[ (1+\delta)^{1/\alpha} -1 \right] +\epsilon,
\label{alf}
\eeq
where the constant $\epsilon$ can be approximated as:
\beq
\epsilon = \frac{\alpha-1}{2\alpha} \sigma_\delta^2
\,.
\label{eps}
\eeq
Formula (\ref{eps}) is accurate to the second order,
we have tested its relevance by fitting 
both $\alpha$ and $\epsilon$ as independent
parameters. The values of $\alpha$ obtained in
these two fits are consistent for filter radii
larger than about 6\hmpc \ (see Figure~\ref{alph}).

We plot the joint PDF of $\de$ and $\te$ combined
from all our six simulations in Figure~\ref{fig:banangs}. 
Willick \& Strauss (1998) performed their VELMOD analysis
of peculiar velocities  using IRAS galaxy density field
convolved with either 3\hmpc\ or 5\hmpc\ Gaussian filter.
They reported very
similar results for both smoothing scales. Therefore, we chose
to plot the distribution for data fitered with 
4\hmpc\ Gaussian kernel.

Formula (\ref{alf}) is drawn in this figure 
with heavy solid line (two-parameter fit), and
dotted line (one-parameter fit, offset accurate to
the second order, formula \ref{eps}). 
The long- and short-dashed lines
represent respectively the third-order polynomial fit
and a formula by Willick \etal (1997):
\beq
\langle\te_{|\de}\rangle = 
      \frac{(1+a^2\sigma_\de^2)\de + a\sigma_\de^2}{1+a\de}
\,,
\label{willick}
\eeq
in which  $a$ is a free parameter.
The polynomial fit 
is poor not only  for very large $\de$, but  it also 
overestimates $\te$ for $1<\de<5$.
The formula of Willick \etal
follows the PDF as closely as ours for 
$ -0.6 < \de < 4 $, but departs from the simulated distribution
at the extreme values of $\de$. 
We have also estimated $a$ from our simulations,
obtaining the value of 0.14 at the 4\hmpc\ scale.
At large scales our estimate of $a$ is rather 0.24
than 0.28 reported by Willick \etal

Our fit for the one-parameter $\alpha$-formula slightly overestimates
$\te$ in the high-$\de$ tail.
%
With the two-parameter
description used instead, 
our formula fits the data very well in the entire range of
$\de$, at a slightly greater $\alpha$.

The weak dependence of $\alpha$ on the smoothing
scale is well visible in Figure~\ref{alph} 
(in the two-parameter description, $\alpha \simeq
1.9 \pm 0.1$). As stated earlier, the formula~(\ref{alf}) as a
description of the inverse relation was our `educated' guess. 
Expanding it 
for large smoothing scales,
and comparing it with the polynomial
description we find that $\alpha \simeq \alpha_{LS} = 1/(1 + 2 r_2)$.
From Figure~\ref{rntg} we obtain $\alpha_{LS}
\simeq 1.95$, which remains in good agreement
with direct fit. For very small smoothing scales, simple
considerations of energy conservation in the model of spherical
collapse yield $\alpha = 2$. 
We expect $\alpha$ to change very weakly 
between weakly and highly non-linear
scales, which is indeed observed.

Our formula is also much simpler than 
formula~(18) of B99 for the forward relation. The reason 
that the formula of B99 is complex is twofold.
Firstly,  B99 aimed at modelling 
the weak $\Omega$-dependence of the relation.
However, the $\Omega$-dependence turned out to be so
weak that it was practically unnecessary to account for it.
Secondly, they  rigorously applied the constraint, coming
from the maximal expansion of voids, that 
$\te = -3/2$ for $\de = -1$.
Although we have not required it explicitely, 
our formula satisfies this `voids' constraint very well
(see Fig.~7).

\section{Summary and conclusions}
\label{sec:suma}

We have tested several parameterizations of the density vs.\
velocity divergence relation on weakly and mildly non-linear
scales. We confirm that the polynomial formula provides a good
description for the forward relation (the density contrast as a
function of the velocity divergence). 

On the other hand, 
the inverse relation is not well described by the
polynomial expansion, which does not converge fast enough. Also, on
mildly non-linear scales the parameters of the expansion strongly
depend on the smoothing scale. The formula of Willick et al.~(1997)
is better than the polynomial description, but it is not free from
drawbacks, either. Firstly, for large densities it has a horizontal 
asymptote, not observed in the simulated distribution. As a
result, in the high-density tail it underestimates the actual
relation. Secondly, like the polynomial description, it incorrectly
describes the low-density tail (i.e., the relation for voids). 

Our formula (7) is free from these disadvantages. It it also simpler (in
its simplest form it depends on one parameter only), which makes it
easy to implement in the velocity-velocity comparisons. 
 
\section*{Acknowledgments}
This work was supported by the KBN grant 2P-03D-004-13. The
simulations are partly performed 
at the {\em Interdisciplinary
Centre for Mathematical and Computational Modelling} in Warsaw.


\end{document}